# Topological and geometrical restrictions, free-boundary problems and self-gravitating fluids.

A. Pelayo[(a)].
Mathematics Department.
University of Michigan.
525 East University Avenue.
48109-1109 MI, Ann Arbor. USA.

D. Peralta-Salas[(b)].
Departamento de Física Teórica II.
Facultad de Ciencias Físicas.
Universidad Complutense.
28040 Madrid. Spain.

## Abstract

Let (P1) be certain elliptic free-boundary problem on a Riemannian manifold $(M,g)$. In this paper we study the restrictions on the topology and geometry of the fibres $f^{-1}(t)$, $t \in f(M)$, of the solutions $f$ to (P1). We give a technique based on certain remarkable property of the fibres (the analytic representation property) for going from the initial PDE to a global analytical characterization of the fibres (the equilibrium partition condition). We study this analytical characterization and obtain several topological and geometrical properties that the fibres of the solutions must possess, depending on the topology of $M$ and the metric tensor $g$. We apply these results to the classical problem in physics of classifying the equilibrium shapes of both Newtonian and relativistic static self-gravitating fluids. We also suggest a relationship with the isometries of a Riemannian manifold.

(a) E-mail address: apelayo@umich.edu. This author is supported by EPSRC and VIGRE grants from the University of Michigan.

(b) E-mail address: dperalta@fis.ucm.es. This author is supported by an FPU grant from Ministerio de Educación, Cultura y Deportes (Spain). Author to whom any correspondence should be addressed.

# 1. Introduction.

In this paper we explore a not very studied property of the solutions to PDE on manifolds: the shapes (the geometry and topology) of the level sets of these solutions. The literature related to this topic [1, 2] is, as far as we know, essentially focused on rather general properties of the level sets; for example, the study of the critical sets of the solutions (sets of vanishing gradient) and the convexity or starshapedness of the level sets. On the contrary we are interested in stronger geometrical and topological characterizations of the level sets: restrictions on the mean and Gauss curvatures, topological types, parallellism, … For general PDE this seems to be a very interesting although a very hard problem since the specific analytical behaviour of the solutions is, in general, unknown. For this reason we restrict ourselves to study certain particular class of PDE on a Riemannian manifold that we will call problem (P1) and whose form is motivated by the equations of static self-gravitating fluids.

More specifically, let $(M,g)$ be an $n$-dimensional Riemannian manifold. We will be concerned with the topological and geometrical restrictions that a solution $f$ of (P1) must have on its fibres $f^{-1}(t)$ with $t \in f(M)$. These restrictions may be a consequence of the existence of intrinsic symmetries but we will not adopt this point of view here. Instead we study these restrictions directly with geometrical rather than with analytical techniques. In this sense we believe that both the approach and the methods used in this paper are new in the literature.

The spaces $M$ for which we obtain more general results not depending on the topology of $M$ are: locally symmetric, conformally flat and constant curvature 3-manifolds. According to Scott [3] five out of the eight canonical 3-dimensional geometries are included in this group of base spaces: $R^3$, $S^3$, $H^3$, $S^2 x R$ and $H^2 x R$ (with their canonical metrics).

Further to the purely mathematical interest, our results are useful in dealing with a problem arising in fluid mechanics: the study of isolated and static self-gravitating fluids. A fluid in these conditions represents a simplified stellar model of fluid-composed star. Depending on whether the gravity force is modelled by Poisson's equation or Einstein's equations we respectively say that the fluid is Newtonian or relativistic. The study of the possible shapes (the forms of the domain $\Omega$ taken up by the fluid) that a fluid can take under these conditions (equilibrium shapes) is a classical problem in fluid mechanics (not yet solved in the relativistic case). The equations involved are a special case of the class of PDE that we study. We refer to the free-boundary problems in the Newtonian and relativistic cases as (P2) and (P3) respectively. Our techniques give a classification of the equilibrium shapes in a more general and geometrical way than the classical approaches do.

In brief the interest of this paper is triple:

(i) The study of the topological and geometrical properties of the fibres of the solutions to certain systems of PDE on a manifold, this being, as far as we know, an unexplored field of research. We relate these properties to the topology and geometry of the base space $(M, g)$. In particular it is very remarkable the analytic representation property across the free-boundary that we prove in section 4. This kind of connections between geometry, topology

     and analytical objects as PDE have always been of high interest in the literature (see sections 4 and 5).

(ii)     We obtain a new kind of mathematical objects that we will call equilibrium functions (see section 3). These functions are of mathematical interest since they have interesting properties related to the geometry and topology of the base space ($M$, $g$) on which they are defined (see sections 5 and 6).

(iii)    The physical application to static self-gravitating fluids. The techniques developed in this paper are completely succesful in Newtonian fluids and give interesting and promising results (although not definitive solutions) in relativistic fluids. These techniques are completely different to the previous ones appeared in the literature of this topic (see section 7). For this reason the authors are optimistic with respect to the approach used in this paper and expect that further research in this line could improve the results that we obtain for relativistic fluids.

The summary follows next. In section 2 we introduce some preliminary results. In section 3 we formulate the problem (P1), and state some previous definitions we need and the fundamental results of the paper. In section 4 we study the property of analytic representation and prove the first main theorem. In section 5 we classify the partitions induced on the ambient space by an equilibrium function defined on it. Section 6 establishes a natural connection between equilibrium functions and the Euclidean group of isometries. In section 7 we study the physical consequences of the theorems proved before in the paper. Finally, in section 8, we state several open problems related to the results we obtain.

## 2. Notation and preliminary results.

In this section we review some general results of real-valued analytic functions and differential geometry. Let $(M, g)$ be an orientable, connected, boundaryless $n$-dimensional Riemannian manifold.

### Real-valued analytic functions.

Consider the set $C^w(M)$ of real-valued analytic functions on $M$.

**Lemma 2.1.** [4] Let $U \subseteq M$ be an open set and let $f : U \to R$ be a real-valued analytic function. Then there always exists an open set $V$ such that $U \subseteq V$ and there is a unique analytic continuation $\tilde{f} : V \to R$ of $f$ to $V$ (that is $\tilde{f}\big|_U = f$ and $\tilde{f}$ is analytic on $V$).

**Corollary 2.2.** If $f$, $g$ are real valued analytic functions on an open set $U \subseteq M$ and there is another open set $W \subseteq U$ such that $f = g$ on $W$ then $f = g$ on $U$.

The following corollary characterises the connected components of the set of critical points of a real-valued analytic function.

**Corollary 2.3.** Let $f$ be a non constant real-valued analytic function on $M$. Then

1. The set of critical points of $f$, $C = \{p \in M : df(P) = 0\}$, has Lebesgue-measure zero.

2. $C$ has no limit sets. In particular if $K \subseteq M$ is compact then $K \cap C$ has a finite number of connected components.

*Proof.* $C$ does not contain any open set $U$ since otherwise the components of the differential of $f$, $df$, would identically vanish on $U$ and hence by corollary 2.2 on $M$.
Let $K$ be a compact set. Let us now see that $C$ has not limit sets and in particular the number of connected components of $K \cap C$ is finite. Should this be otherwise then inside of $K$, $C$ would have a limit set $L$. By continuity $df(L)=0$. $C$ is by definition an analytic set of $M$ [4] whose connected components have codimension $\geq 1$. Now, Lojaciewicz's structure theorem [4] claims that a real analytic set can be stratified into a discrete set of submanifolds of dimension 0, 1, ..., $n$-1. This property immediately implies the non-existence of limit sets $L$. □

Fixing the notation of $C$ for the set of critical points of a real-valued analytic function it is straightforward to prove the following corollary from corollary 2.3.

**Corollary 2.4.** If $f$ is analytic on $M$ then the set of critical values $f(C)$ is a countable set of isolated points.

For our purposes a submanifold $S$ of $M$ is analytic if it can be realised globally as the zero-set, $\{p \in M : f(p) = 0\}$, of a real-valued analytic function $f$ on $M$.

Let $f$ be a continuous function on $M$. For each $c \in f(M)$ we have a preimage $f^{-1}(c)$. Let $V_c^i$ for $i \in I(c)$ be the connected components of $f^{-1}(c)$ (where $I(c) = \{1,2,...,$ number of connected components of $f^{-1}(c)\}$). We denote by $\beta_M(f)$ the following set:

$$\beta_M(f) = \bigcup_{\substack{c \in f(M) \\ i \in I(c)}} V_c^i \quad . \tag{2.1}$$

For an arbitrary continuous map $f$ we will say that the set $\beta_M(f)$ is the partition induced on $M$ by $f$. We usually do not mention the function $f$ determining a partition $\mathcal{A}$ on a manifold $M$, but we just say that $\mathcal{A}$ is a partition of $M$. If we have an open set $U \subset M$ the partition induced on $U$ by $f|_U$ is called $\beta_U(f)$.

**Definition 2.5.** If $U \subset M$ is an open set such that it is constituted of connected components of fibres of $f$ and hence $U = \bigcup_{\substack{c \in B \subset f(M) \\ i \in J(c) \subset I(c)}} V_c^i$ we say that $U$ is an $f$-partitioned open set.

**Definition 2.6.** Let $U \subset M$ be an $f$-partitioned open set. We say that a partition $\beta_U(f)$ is locally trivial across a fibre $F \subset U$ of $f$ if for the point $f(F) \subset f(U)$ there exists an open neighbourhood $I \subseteq f(U)$ and a homeomorphism $h : I \times F \to f^{-1}(I) \subseteq U$ such that for each $t \in I$ the correspondence $x \to h(t,x)$ defines a homeomorphism between $F$ and the fibres $f^{-1}(t)$ in $U$. If the partition $\mathcal{A}$ of $U$ is locally trivial across every fibre $F \subset U$ then we say that $\mathcal{A}$ is locally trivial on $U$.

With this in mind we do the following important definitions.

**Definition 2.7.** Let $\mathcal{A}$ be a partition of an open set $U \subset M$. A function $f$ represents the partition $\mathcal{A}$ on $U$ if $\mathcal{A} = \beta_U(f)$. We say that $\mathcal{A}$ admits an analytic representation if there exists a real-valued analytic function $f$ on $U$ such that $\mathcal{A} = \beta_U(f)$.

**Definition 2.8.** We say that a function $f$ is analytically representable on the open set $U$ if $\beta_U(f)$ admits an analytic representation.

We will say that $f: M \to R$ agrees fibrewise with $g: M \to R$ if $\beta_M(f) = \beta_M(g)$. Naturally $f_1, \ldots, f_n$ agree fibrewise if for all $i,j$ $f_i, f_j$ agree fibrewise.

**Lemma 2.9.** Let $f, g$ be real-valued analytic functions on $M$. Let $U \subset M$ be an open set. Then $\beta_U(f) = \beta_U(g) \Rightarrow \beta_M(f) = \beta_M(g)$.

*Proof.* Since $\beta_U(f) = \beta_U(g)$ we have that

$$rk \begin{pmatrix} df \\ dg \end{pmatrix} \leq 1 \quad \text{in } U \quad . \tag{2.2}$$

Since $U$ is an open set and $f$, $g$ are analytic functions we have that condition (2.2) is satisfied not only in $U$ but in the whole $M$

$$rk\begin{pmatrix} df \\ dg \end{pmatrix} \leq 1 \quad \text{in } M \quad . \tag{2.3}$$

Condition (2.3) implies [5] that $f$ and $g$ are functionally dependent and so there exists (at least locally) an analytic function $Q: R^2 \to R$ such that

$$Q(f,g) = 0 \quad \text{in } M \quad . \tag{2.4}$$

Equation (2.4) shows that the partitions induced on $M$ by $f$ and $g$ must agree. □

## Riemannian Geometry [6].

Unless otherwise stated the metric $g$ is analytic, the manifold $M$ is also analytic and the Riemannian space $(M, g)$ is complete. A repetition of a parameter as a subscript and superscript in the same expression means that we sum in that index.

We denote the set of analytic vector fields on $M$ by $T^w(M)$. As usual the terms $\nabla$, $\Delta$ and $div$ will stand for the standard gradient, Laplace-Beltrami and divergence operators on a Riemannian manifold.

In a local coordinate system the Riemann curvature tensor, the Ricci tensor and the scalar curvature of the manifold $(M,g)$ are given by $R^a_{bcd}$, $R_{bd} = R^a_{bad}$ and $R = R^a_a$ respectively. We denote the covariant derivative of a tensor field $T^{a_1..a_n}_{b_1..b_n}$ by $T^{a_1..a_n}_{b_1..b_n;c}$ and the covariant derivative along $X \in T^w(M)$ by $D_X T$.

Let $S$ be a codimension 1 submanifold in $M$ (or regular hypersurface). The metric induced by $g$ on $S$ is given by $\boldsymbol{b}_{ab} = g_{ab} - n_a n_b$, where $n^a$ is the unit normal vector field to $S$ on $M$. As usual $\boldsymbol{b}_{ab}$ defines a covariant derivative on $S$ that we denote by the subscript $\|$. By convention $\tilde{n}^a$ will mean the extension of $n^a$ to a neighbourhood of $S$ in $M$ and the same convention applies for more general tensor fields on S.

The extrinsic curvature or second fundamental form of $S$ is given by

$$H_{ab} = \boldsymbol{b}^c_a \boldsymbol{b}^d_b \tilde{n}_{d;c} = \frac{1}{2} L_{\tilde{n}}\left(\tilde{\boldsymbol{b}}_{ab}\right) \quad , \tag{2.5}$$

$L_{\tilde{n}}$ standing for the Lie derivative respect to the unit normal.

We denote the Riemann curvature tensor induced by $R^a_{bcd}$ on $S$ (with the induced metric) by $R'^a_{bcd}$. It is readily obtained the following equation from Gauss theorem:

$$R' = R - 2R_{ab}n^a n^b + (H^a_a)^2 - H_{ab}H^{ab} \quad . \tag{2.6}$$

From (2.6) we deduce the following relation between the intrinsic sectional curvature of $S$ ($K'$), the sectional curvature of $M$ restricted to $S$ ($K$) and the Gauss curvature of $S$ ($\bar{K}$)

$$K' = K + \bar{K} \quad . \tag{2.7}$$

The sectional curvature of $M$ spanned by two linearly independent vectors $u, v \in T_p M$ at the point $p$ is given by:

$$K = \frac{R_{abcd} u^a v^b u^c v^d}{(g_{ac} g_{bd} - g_{ad} g_{bc}) u^a v^b u^c v^d} \quad . \tag{2.8}$$

We conclude with two remarks. The first is that if $S$ is expressed as the zero set of a smooth function $f$, $S = \{p \in M : f(p) = 0\}$ with $df|_S \neq 0$, then the mean curvature $H$ is the divergence of the unit normal vector field to $S$:

$$H = H_a^a \text{ (definition)} = b_a^c b_d^a \tilde{n}_{;c}^d = div\left(\frac{\nabla f}{\|\nabla f\|}\right). \tag{2.9}$$

The second one is a lemma which characterizes the topology of the connected components of the fibres of a smooth function.

**Lemma 2.10.** Let $f$ be a smooth function on an $f$-partitioned open set $U \subset M$. If $\|\nabla f\| \geq m > 0$ on $U$ then the connected components of $f^{-1}(t)$ in $U$ are diffeomorphic to the connected components of $f^{-1}(t')$ in $U$ for all $t, t' \in f(U)$ and $b_U(f)$ is a locally trivial partition.

*Proof.* The normal vector field $X = \dfrac{\nabla f}{\|\nabla f\|^2}$ is smooth on $U$ and a symmetry of $f$ because $L_X f = 1$ [7]. $X$ defines a uniparametric local group wherever $X \neq 0$ and in fact it also defines a uniparametric global group on $U$ (that is, $X$ is a complete vector field) since $\|X\|$ is bounded above ($\|X\| \leq \dfrac{1}{m}$) [8] and the space is complete. $X$ is always transitive (in appropriate coordinates we can set $X = \partial_f$). Since the flow is also a diffeomorphism the proof is complete. □

# 3. Formulation of the problem and statement of the main theorems.

The problem (P1) we are interested in is a system of partial differential equations defined on $M$. Its form and additional regularity assumptions are inspired by the equations modeling static self-gravitating fluids. For the rest of the paper and unless otherwise stated $\Omega \subset M$ is a connected open set with boundary a codimension one analytic submanifold $\partial\Omega$ (connected or not connected). $f_2, f_3 : M \to R$ are functions with support the open set $\Omega \subset M$ in which they are analytic and constant in $\partial\Omega$. Besides $f_3$ is not identically constant on $\Omega$ and $f_2$ is not identically zero on $\Omega$. With these assumptions in mind we state now the equations of problem (P1):

$$\left.\begin{aligned}
\Delta f_1 &= F(f_1, f_2, f_3) \quad \text{in } \Omega \\
H(f_1)\nabla f_3 &+ G(f_2, f_3)\nabla f_1 = 0 \quad \text{in } \Omega \\
f_1 &= c \ (c \in R), \nabla f_1 \neq 0 \text{ and } f_1 \in C_t^2 \quad \text{on } \partial\Omega \\
\Delta f_1 &= 0 \quad \text{in } M - \overline{\Omega}
\end{aligned}\right\} \qquad (3.1\text{-}3.4)$$

where $F \in C^{\omega}(R^3)$, $G \in C^{\omega}(R^2)$ and $H \in C^{\omega}(R)$ are not identically zero real-valued functions. We also impose $G_{,f_2} \neq 0$ and $G_{,f_3} \neq 0$ in their respective domains. The symbol $C_t^2$ means that $f_1$ is $C^1$ on $\partial\Omega$ and the tangential components of the second derivatives are also continuous, that is $f_{1,ij}t^i$, $j = 1,...,n$, are continuous for any tangent vector field $t = t^i \partial_i$ to the free-boundary. This assumption is analogous to the junction condition of Synge for the metric tensor in general relativity [23]. The normal components of the second derivatives will not be continuous in general.

Note that by the assumption $\nabla f_1 \neq 0$ on $\partial\Omega$ we mean that the gradient is not identically zero on the boundary, but it could be zero on a nowhere dense subset of it (all the theorems of this work hold in this case). Note also that it is only necessary to require that $\partial\Omega$ be smooth enough, its analyticity follows from this assumption and the properties of elliptic free-boundary equations [9].

This kind of problems where we do not specify a priori the domain $\Omega$ but only certain regularity conditions as equation (3.3) are called free-boundary problems (note that the constant $c$ is not a priori prescribed). The unknowns in the problem (P1) are the domain $\Omega$ itself and the functions $(f_1, f_2, f_3)$ that must satisfy not only PDE appearing in (3.1), (3.2) and (3.4) but the regularity conditions stated above.

Note the reader that for certain values of $F$, $G$ and $H$ problem (P1) could not have any solution. Since we are not interested in the existence problem we will suppose that solutions exist and will obtain the restrictions that the fibres of these solutions must possess. The same applies to the base manifold $(M,g)$.

Equations (3.1) and (3.4) are elliptic because so is the Laplace-Beltrami operator. The following lemma is immediate to prove (see reference [10]).

**Lemma 3.1.** The solutions $f_1: M \to R$ to (P1) are analytic on $M - \partial\Omega$.

We next check that equation (3.2) implies that $f_1, f_2$ and $f_3$ agree fibrewise in $\Omega$.

**Lemma 3.2.** The analytic functions $f_1, f_2, f_3$ satisfying equation (3.2) agree fibrewise.

*Proof.* From (3.2) it follows that $d\left(\dfrac{df_3}{G(f_2,f_3)}\right) = -d\left(\dfrac{df_1}{H(f_1)}\right) = 0$ and hence $G_{,f_2} df_2 \wedge df_3 = 0$. Therefore $\nabla f_2$ and $\nabla f_3$ are linearly dependent. Also from (3.2) $df_3 = -\dfrac{G(f_2,f_3)}{H(f_1)} df_1$, taking the exterior derivative in this equation we get $\left(G_{,f_2} df_2 + G_{,f_3} df_3\right) \wedge df_1 = 0$. The linear dependence of $\nabla f_2$ and $\nabla f_3$ implies the linear dependence of all $\nabla f_1, \nabla f_2$ and $\nabla f_3$. This fact and the analiticity of the functions (see section 2) imply that they agree fibrewise in the whole $\Omega$. □

As we explained in section 1 in this paper we are interested in the topological and geometrical restrictions that the problem (P1) forces in the fibres of a solution $f \equiv (f_1, f_2, f_3)$. By lemma 3.2 $b_\Omega(f_1) = b_\Omega(f_2) = b_\Omega(f_3)$ and hence we will reduce our study to look only at the fibres of $f_1$ (which are defined on the whole $M$) that we will call the fibres of $f$ (we will also use the term $b_M(f)$ refering to $b_M(f_1)$).

**Definition 3.3.** A real-valued analytic function $f$ on $M$ is an equilibrium function on $U \subseteq M$ if $f, \|\nabla f\|^2$ and $\Delta f$ agree fibrewise on $U$. Similarly a partition $\acute{A}$ of $U \subseteq M$ is an equilibrium partition of $U$ if there is an equilibrium function $f$ such that $\acute{A} = b_U(f)$. When we do not specify the open set $U$ we assume that the equilibrium function condition holds in the whole $M$.

**Definition 3.4.** A partition $b_M(f)$ of $M$ is said to have a fibre bundle local structure $b_M^*(f)$ if there exists a countable collection of $f$-partitioned open sets $U_i \subseteq M$, $M = \bigcup_i \overline{U}_i$, such that $b_{U_i}^*(f)$ is a rank $(n-1)$ fibre bundle for every $i=1,2,...$ and the fibre of $b_{U_i}^*(f)$ is a regular connected submanifold of $M$ for every $i=1,2,....$ We say that the fibres of $b_M^*(f)$ are the regular connected codimension 1 submanifolds of $b_M(f)$.

If $b \in [0, \infty)$ we denote by $bS^p = \{x \in (R^{p+1}, d) : \|x\| = b\}$ with its standard Riemannian metric. Taking cartesian product we obtain a new Riemannian manifold $bS^p \times R^q$ in $(R^{p+q+1}, d)$ (with the standard product metric). We will call the elements $bS^p \times R^q$ as $b, p, q$ vary in their respectives domains, standard cylinders.

**Definition 3.5.** A partition $\acute{A}$ is said to be almost-trivial if it has a fibre bundle local structure $\acute{A}^*$ and every fibre of $\acute{A}^*$ has constant principal curvatures and is geodesically parallel to its neighbour fibres. If furthermore all the fibres are (globally) isometric to a standard cylinder $bS^p \times R^q$ with $p, q$ fixed natural numbers, then $\acute{A}$ is said to be geometrically trivial.

The following three are the main theorems of this paper.

**Theorem 3.6.** If $f$ is a solution of (P1) then $\boldsymbol{b}_M(f)$ is an equilibrium partition of $M$.

**Theorem 3.7.** Let $f$ be an equilibrium function on a Riemannian 3-manifold $M$ satisfying either of the following

1. $M$ is flat.
2. $M$ is conformally flat and $f$, $\boldsymbol{f}$ agree fibrewise where $g_{ab} = e^{2f}\boldsymbol{d}_{ab}$.
3. $M$ is locally symmetric and each regular connected component of $f^{-1}(t)$ ($t\boldsymbol{\hat{I}}f(M)$) has parallel second fundamental form.

Then $\boldsymbol{b}_M(f)$ is almost-trivial.

**Theorem 3.8.** Let $f$ be an equilibrium function on a Riemannian $n$-manifold $M$ which is diffeomorphic to $R^n$ and satisfying either of the following

1. $M$ is flat.
2. $M$ is conformally flat and $f$, $\boldsymbol{f}$ agree fibrewise where $g_{ab} = e^{2f}\boldsymbol{d}_{ab}$.

Then $\boldsymbol{b}_M(f)$ is geometrically trivial.

**Corollary 3.9.** If $M$ is like in theorem 3.7 or theorem 3.8 and $f$ is a solution of (P1) then $\boldsymbol{b}_M(f)$ is almost-trivial or geometrically trivial respectively.

*Proof.* It is straightforward from theorems 3.6, 3.7 and 3.8. □

Note that the classification obtained in corollary 3.9 is only a necessary condition on the partitions induced by the solutions. It is not a sufficient condition in the sense that any $f$ verifying that $\boldsymbol{b}_M(f)$ is geometrically trivial or almost-trivial need not be a solution of (P1). There will be analytical restrictions (that could prevent the existence of solutions for certain $F$, $G$, $H$ and $(M,g)$) but we will not study them here.

# 4. Analytic representation and proof of theorem 3.6.

As it is shown in Lemma 3.1 the solutions $f_1$ to problem (P1) are analytic on the whole $M$ except for the boundary $\partial\Omega$ where we only require to be $C^1$. This fact implies that the existence of an analytic representation of the partition induced by $f$ on a neighbourhood of $\partial\Omega$ is not immediate nor trivial. We are interested in studying the partitions induced on $M$ by $f$ (solutions to (P1)). From this point of view it is reasonable to try to represent those partitions by functions as good as possible, that is, analytic functions. Therefore, the philosophy of an analytic representation technique consists of substituting the pathological function $g$ representing a partition on $U \subset M$ by an analytic function $f$ representing the same partition ($\boldsymbol{b}_U(g) = \boldsymbol{b}_U(f)$). This is the best approach we have found for working analytically with partitions (essentially topological and geometrical objects) underlying a system of PDE.

We next study when the fibres of a $C^k$ ($k \geq 1$) function defined on $M$ can be realised as fibres of an analytic function in a neighbourhood of certain fibre.

**Definition 4.1.** We say that a real-valued analytic function $f: M \to R$ is an analytic representation of the fibres of $g: M \to R$ on a neighbourhood $U$ of certain connected component of $g^{-1}(t)$ ($t \in g(M)$) if $\boldsymbol{b}_U(g) = \boldsymbol{b}_U(f)$.

The following example illustrates the fact that given a non analytic function it is reasonable to look for an analytic function agreeing fibrewise with it.

**Example 4.2.** Let $f: R^n \to R$ be given by $f(x) = \|x\|^2 - 1$ ($\|\ \|$ standing for the Euclidean norm) and $h: R \to R$, $h(t) = te^{-1/t^2}$ if $t \neq 0$ and $0$ otherwise. Then the function $g = f + h \circ f$ is $C^\infty$ on $R^n$, analytic on $R^n - g^{-1}(0)$ and agrees fibrewise with $f$. So $f$ is an analytic representation of the fibres of $g = f + h \circ f$ in a neighbourhood of $g^{-1}(0)$.

Nevertheless there are functions which do not admit an analytic representation in any neighbourhood of some of the connected components of their fibres. The following examples with non-compact and compact fibres illustrate this fact.

**Example 4.3.** Let $g: R^n \to R$ be given by (in cartesian coordinates)

$$g(x^1,...,x^n) = \begin{cases} x^1\left(1 + x^2 e^{-1/(x^1)^2}\right) & \text{if } x^1 > 0 \\ x^1 & \text{if } x^1 \leq 0 \end{cases}. \qquad (4.1)$$

We claim that there is no an analytic representation $f$ of the fibres of $g$ in any neighbourhood of $g^{-1}(0)$. Define $\boldsymbol{f}: R \to R$ by $u \to \boldsymbol{f}(u)$ where $f(\boldsymbol{f}(u),1,0,...,0) = f(u,-1,0,...,0)$ and $(\boldsymbol{f}(u),1,0,...,0)$ lies in the same connected component of the fibre as $(u,-1,0,...,0)$. Since $f$ is analytic, by the implicit function theorem so is $\boldsymbol{f}$ but since $\boldsymbol{f}(u) = u$ when $x^1 \leq 0$ and different otherwise $\boldsymbol{f}$ is not analytic.

**Example 4.4.** Let $g: R^2 \to R$ be given (in polar coordinates) by

$$g(r,\mathbf{q}) = \begin{cases} 1 - r^2 + \sin\mathbf{q}\, e^{-1/(1-r)^2} & \text{if } r \geq 1 \\ 1 - r^2 & \text{if } r < 1 \end{cases}. \qquad (4.2)$$

By the same argument as in the previous example, we conclude that there is no analytic representation of $g$ in any neighbourhood of $g^{-1}(0)$.

*Important remark.* Looking at these examples we realize what is happening near $g^{-1}(0)$ (which we assume connected) and can proceed with a general argument. Let $g: M \to R$ be a $C^k$ function and assume that $dg|_{g^{-1}(0)} \neq 0$. Say $\frac{\partial g}{\partial x^1} \neq 0$ in local coordinates $(x^1,...,x^n)$ in certain neighbourhood $U$ of a point $p \in g^{-1}(0)$. Let $\mathbf{g}: (-\mathbf{d},\mathbf{d}) \to U \subset M$ be a $C^k$ curve transversal to $g^{-1}(0)$ at the fixed point $\mathbf{g}(0) = p \in g^{-1}(0)$. We will say that $g$ is a candidate to be analytically representable if $g \circ \mathbf{g}: R \to R$ is analytic. This definition does not depend on the choice of $\gamma$ and if we call $\mathbf{f} \equiv (g \circ \mathbf{g})^{-1}$ it is clear that $\mathbf{f} \circ g$ and $g$ agree fibrewise in a small open neighbourhood of $g^{-1}(0)$. So if $\mathbf{f} \circ g$ is analytic we are done. If $\mathbf{f} \circ g$ is not analytic then it is enough to notice that in general if $i,j: M \to R$ are $C^k$ functions agreeing fibrewise and candidate to be analytically representable, then there is a unique analytic diffeomorphism $\mathbf{t}: R \to R$ such that in $U$ $i = \mathbf{t} \circ j$ and it is given by the formula $\mathbf{t} = (i \circ \mathbf{g}) \circ (j \circ \mathbf{g})^{-1}$. Hence we have proved that the map $g \circ \mathbf{g}$ is a diffeomorphism with inverse $\mathbf{f} \equiv (g \circ \mathbf{g})^{-1}$. Further, if $\mathbf{f} \circ g$ is not analytic then there is no possible analytic representation of the fibres of $g$ in any neighbourhood of $g^{-1}(0)$.

All these results show that it is not evident that any solution $f$ of the problem (P1) be analytically representable on a neighbourhood of $\partial\Omega$. Anyway there are physical reasons which suggest that the fibres of the solutions should be analytically representable across the free-boundary (recall that (P1) generalizes the physical problems (P2) and (P3)). On the one hand the results of Broglia and Tognoli [11] ensure the existence of a partition admitting analytic representation near the pathological, but physically two arbitrarily close partitions are indistinguishable. On the other hand it would be rather surprising that examples like 4.3 and 4.4 could represent the equilibrium equipotential hypersurfaces of static self-gravitating fluids.

Indeed the problem (P1) has the remarkable property of analytic representation, which we will prove at the end of this section.

**Theorem 4.5: analytic representation property (ARP).** The solutions $f$ of the problem (P1) are analytically representable on a neighbourhood of $\partial\Omega$ by functions $I$ analytic on the whole $M$.

Concerning the physical meaning of ARP we must say the following. If the matched function $f$ were not analytically representable in a neighbourhood of $\partial\Omega$ we would have that the matching would be smooth but however the inner region would be independent of the outer region (see examples 4.3 and 4.4). Therefore ARP means that a physical matching (at least in static situations) must not only guarantee the continuity of the gravitatory field but also the physical dependence between the external and internal properties of the fluid.

Recall that in the literature concerning the physical problems (P2) and (P3) (particular cases of (P1)) <u>strong</u> assumptions are imposed in order to characterize the geometry of the equipotential surfaces. For instance, $\|\nabla V\|$ is a function of the potential $V$ only [29], there exists a "reference spherical model" [24], the existence of equation of state and asymptotic conditions [25] … see section 7 for more details. In our work no additional assumptions are necessary.

*Proof of theorem 3.6.* We fix our attention in a connected component $(\partial\Omega)_r$ of $\partial\Omega$. In fact, we are going to prove that there exists an open set $U$ in a neighbourhood $L$ of $(\partial\Omega)_r$ ($L \subset \Omega$) where $f_2$ and $f_3$ are functions of $f_1$. We suppose that $\nabla f_1$ does not vanish in $L$ (it is always possible by continuity if $L$ is small enough). For example if $f_{1,x^1} \neq 0$ in $U$ (we are considering a local coordinate system $(x^1, \ldots, x^n)$) the implicit function theorem guarantees the following steps:

$$x^1 = f_1^{-1}(f_1, x^2, \ldots, x^n) \Rightarrow f_2 = f_2(f_1^{-1}(f_1, x^2, \ldots, x^n), x^2, \ldots, x^n) = \tilde{f}_2(f_1, x^2, \ldots, x^n)$$
$$f_3 = f_3(f_1^{-1}(f_1, x^2, \ldots, x^n), x^2, \ldots, x^n) = \tilde{f}_3(f_1, x^2, \ldots, x^n) \ .$$

It is easy to check that $\tilde{f}_{2,x^2} = \ldots = \tilde{f}_{2,x^n} = \tilde{f}_{3,x^2} = \ldots = \tilde{f}_{3,x^n} = 0$. One only has to take into account the implicit function theorem and that $f_1, f_2, f_3$ agree fibrewise. Hence in $U$ $f_2 = \tilde{f}_2(f_1), f_3 = \tilde{f}_3(f_1)$ where $\tilde{f}_2$ and $\tilde{f}_3$ are analytic functions of its argument. Therefore by equation (3.1) $\Delta f_1 = \tilde{F}(f_1)$ in $U$ ($\tilde{F}$ analytic). In the outer region $\Delta f_1$ vanishes identically.

On account of ARP $b_M(f_1)$ admits an analytic representation in a neighbourhood $S$ of $(\partial\Omega)_r$ by a new function $I$ analytic on $M$, $b_\Sigma(f_1) = b_\Sigma(I)$, in fact the analiticity of $f_1$ in $M$ - $\partial\Omega$ implies that $b_M(f_1) = b_M(I)$ in the whole $M$ (see lemma 2.9). Let $G \subset S$ be an open neighbourhood of $(\partial\Omega)_r$ whose intersection with $\Omega$ and $M - \bar{\Omega}$ is not the empty set. Again, by the implicit function theorem, we can express $f_1$ as a function of $I$ in certain open set $V$, $f_1 = R(I)$, $R$ standing for an analytic function on $V \subset G$ except in $V \cap (\partial\Omega)_r$ where it is only $C_t^2$ (suppose that $V \cap \Omega \neq \emptyset, V \cap (M - \bar{\Omega}) \neq \emptyset$). Now define

$$V_{\text{in}} = V \cap \Omega \subset U \quad \text{and} \quad V_{\text{out}} = V \cap (M - \bar{\Omega}) \tag{4.3}$$

From equation (3.4) it follows $\Delta R(I) = 0$ in $V_{\text{out}}$. But $\Delta R(I) = 0$ implies, since $R'(I) \neq 0$ in $V$ (because $\nabla f_1 \neq 0$ if $G$ is small enough) that

$$R''(I)\|\nabla I\|^2 + R'(I)\Delta I = 0 \quad , \tag{4.4}$$

i.e. $\dfrac{\Delta I}{\|\nabla I\|^2} = -\dfrac{R''(I)}{R'(I)}$. Therefore $\dfrac{\Delta I}{\|\nabla I\|^2} = c(I)$ in the outer region $V_{\text{out}}$. Since $I$ is analytic so are $\Delta I$ and $\|\nabla I\|^2$ (the metric tensor is analytic) and hence $\dfrac{\Delta I}{\|\nabla I\|^2}$ wherever $\nabla I \neq 0$ and in particular in $V$. Let $\tilde{c}$ be the analytic continuation of $c$ to $V$ (this continuation indeed exists because of the analiticity of $I$ and $\dfrac{\Delta I}{\|\nabla I\|^2}$ on $V$). Certainly $\tilde{c}(I) = \dfrac{\Delta I}{\|\nabla I\|^2}$ in $V$ and in particular in $V_{\text{in}}$. On the other hand

$$R''(I)\|\nabla I\|^2 + R'(I)\Delta I = \tilde{\tilde{F}}(I) \tag{4.5}$$

in $V_{\text{in}}$ and together with (4.4) implies that $\Delta I$ and $\|\nabla I\|^2$ depend only on $I$. The argument applies to the whole of $V$ by the analiticity of $\Delta I$ and $\|\nabla I\|^2$, i.e. $\Delta I = y(I)$ and $\|\nabla I\|^2 = w(I)$ in $V$. Again because of the analiticity of all of $\Delta I$, $\|\nabla I\|^2$ and $I$ the property of equilibrium partition extends to $M$ (although generally we can write one as a function of the other only locally). Since $I$, $f_1$ agree fibrewise we have that $b_M(f_1)$ is an equilibrium partition. □

*Proof of theorem 4.5.* This proof is inspired by the remarkable technique developed by Lindblom in [30]. In the open set $V$ we can write the equations defining (P1) in terms of just $f_1$:

$$\left. \begin{array}{l} \Delta f_1 = \tilde{F}(f_1) \text{ in } V_{\text{in}} \\ f_1 = c, c \in R, \nabla f_1 \neq 0 \text{ and } f_1 \in C_t^2 \text{ on } \partial\Omega \cap V \\ \Delta f_1 = 0 \text{ in } V_{\text{out}} \end{array} \right\} \tag{4.6}$$

Consider the Lie algebra $L$ of symmetries of $f_1$ in $V_{\text{in}}$ and take a generic vector field $\boldsymbol{x} \in L$, that is $\boldsymbol{x}(f_1) = 0$. Note that $\boldsymbol{x} = \boldsymbol{x}^i \partial_i$ can always be chosen analytic. Since the free-boundary is analytic then this analytic symmetry can be extended to $\partial\Omega$ in such a way that $(\boldsymbol{x}^i)_{|\partial\Omega} \in C^w$ and $(\boldsymbol{x}^i_{,j})_{|\partial\Omega} \in C^w$. Now we are going to extend the vector field $\boldsymbol{x}$ beyond the free-boundary $\partial\Omega$ such that the extension be analytic in the whole $V$. $\boldsymbol{x}$ in $V_{\text{out}}$ is defined by the following problem:

$$\Delta \boldsymbol{x}^k + \dfrac{2D_m D_k f_1}{f_{1,k}} \left( D^m \boldsymbol{x}^k \right) + \dfrac{R_{ik} D^i f_1}{f_{1,k}} \boldsymbol{x}^k = 0 \tag{4.7}$$

$k = 1,\ldots,n$, provided with the boundary conditions on $\partial\Omega$ given by the values $(\boldsymbol{x}^i)_{|\partial\Omega}$ and $(\boldsymbol{x}^i_{,j})_{|\partial\Omega}$ of the interior symmetry. Note that the symbol $D$ stands for the covariant derivative. In order that (4.7) be well defined assume, without loss of generality, that in $V$ all the components of $\nabla f_1$ are non-zero. Since all the terms of (4.7) are analytic, the

boundary conditions and the free-boundary are also analytic and the equation is elliptic (and linear) then there exists a unique extension which is an analytic vector field in $V_{\text{out}}$ [10] and therefore we have constructed an analytic $\mathbf{x}$ in $V$.

The extended vector field $\mathbf{x}$ has the remarkable property of being a symmetry of $f_1$ in $V_{\text{out}}$. Indeed, the first step consists of the following computation

$$\Delta(\mathbf{x}^k f_{1,k}) = D_m D^m (\mathbf{x}^k f_{1,k}) = f_{1,k}\left(\Delta \mathbf{x}^k + \frac{2 D_m D_k f_1}{f_{1,k}}(D^m \mathbf{x}^k) + \frac{D^m D_m D_k f_1}{f_{1,k}} \mathbf{x}^k\right) \qquad (4.8)$$

Note now that $D_m D^m D_k f_1 = D_k(\Delta f_1) + R_{ik} D^i f_1$ and since the first term of this equation is zero ($f_1$ is solution of (P1)) we get from equations (4.7) and (4.8) that

$$\Delta(\mathbf{x}^k f_{1,k}) = 0 \qquad (4.9)$$

in $V_{\text{out}}$. In this case the boundary conditions are $\left(f_{1,k}\mathbf{x}^k\right)_{|\partial\Omega} = 0$ since $f_1$ is $C^1$ on the boundary, and $\partial_i\left(f_{1,k}\mathbf{x}^k\right)_{|\partial\Omega} = \left(f_{1,ij}\mathbf{x}^j + f_{1,j}\mathbf{x}^j_{,i}\right)_{|\partial\Omega} = 0$ because the second derivatives of $f_1$ are continuous in the tangential direction to the boundary and $\mathbf{x}$ is tangent to $\partial\Omega$. Cauchy-Kowalewsky theorem for elliptic equations [10] implies that the solution to (4.9) is $\mathbf{x}^k f_{1,k} = 0$ in $V_{\text{out}}$. The globalization to the whole of $M$ immediately follows from the analyticity of $f_1$ in $M - \partial\Omega$.

Summarizing we have a Lie algebra $L$ of $n$-1 independent (up to a null measure set) analytic symmetries of $f_1$ in $M$. From this Lie algebra we can reconstruct the partition $\mathbf{b}_M(f_1)$ via Frobenius theorem [19]. Since the vector fields of $L$ are analytic it is ready to construct a vector field normal to the components of $\mathbf{b}_V(f_1)$ which is analytic as well and therefore we get an analytic curve transversal to the partition (across the free-boundary). On account of the necessary and sufficient condition proved in the preceding remark we conclude that an analytic reconstruction and therefore an analytic representation $I$ of $\mathbf{b}_M(f_1)$ in $M$ is possible. $\square$

# 5. The classification theorems.

In section 4 we have related the solutions of (P1) to the equilibrium functions on a Riemannian manifold (*M*,*g*) (theorem 3.6). The problem of classifying the partitions induced by the solutions of (P1) has been substituted by the problem of classifying the equilibrium partitions on different spaces. We have reduced the original problem involving a difficult system of PDE to a purely geometrical problem. In this section we give several topological and geometrical properties of the equilibrium partitions on different Riemannian manifolds.

## General results.

Firstly we give a theorem characterizing certain general properties that all the equilibrium partitions must possess on any Riemannian manifold.

**Theorem 5.1.** Let (*M*,*g*) be an *n*-dimensional Riemannian manifold and *f* an equilibrium function on *M*. Then the partition induced by *f* on *M*, $\boldsymbol{b}_M(f)$, has a fibre bundle local structure $\boldsymbol{b}_M^*(f)$, each fibre of $\boldsymbol{b}_M^*(f)$ has constant mean curvature and locally the fibres of $\boldsymbol{b}_M^*(f)$ are geodesically parallel.

*Proof.* Since *f* is analytic the number of critical values *f(C)* of *f* is countable (corollary 2.4). Without loss of generality suppose that $f(M)=(-\infty,+\infty)$ and that there are only *N* critical values, say $c_1 < c_2 < ... < c_N$ ($c_i \in R$) and hence *M* is divided into disconnected open regions $M_1 = f^{-1}((-\infty \equiv c_0, c_1))$, $M_{i+1} = f^{-1}((c_i, c_{i+1}))$ and $M_{N+1} = f^{-1}((c_N, +\infty \equiv c_{N+1}))$ $\Rightarrow M = \bigcup_{i=1}^{N+1} M_i \cup C(f)$. Note that each $M_i$ may be made up by several connected components $M_i^j$, $j \hat{I} J(i)$ ($J(i) = \{1,2,...,$ number of connected components of $M_i\}$). Take *e*>0, the equilibrium condition implies that $\|\nabla f\| \geq m > 0$ on the *f*-partitioned open sets $f^{-1}((-\frac{1}{e}, c_1))$, $f^{-1}((c_i + e, c_{i+1} - e))$, $f^{-1}((c_N, \frac{1}{e}))$ and hence by lemma 2.10 and doing *e* tend to zero we have that the connected components of the fibres of *f* on each $M_i^j$ are diffeomorphic each other and the partitions $\boldsymbol{b}_{M_i^j}(f)$ are locally trivial. Note that the hypersurfaces on different connected components of $M_i$ or on $M_i$ and $M_j$ ($i \neq j$) need not be in general diffeomorphic between them due to the existence of critical fibres $f^{-1}(c_i)$ of separation.

Denote the set of regular values of *f* by $R(f) = f(M) - f(C)$. We construct a fibre bundle local structure $\boldsymbol{b}_M^*(f)$ over *R(f)* by the following. The bundle space is given by $\boldsymbol{b}_{M_i^j}(f)$ and the projection map $\boldsymbol{p}_M(f) : \boldsymbol{b}_{M_i^j}(f) \to (c_i, c_{i+1}) \subset R(f)$ naturally by: the fibre over $y \in (c_i, c_{i+1})$ is the connected component of $f^{-1}(y)$ in $M_i^j$. Clearly $\boldsymbol{b}_{M_i^j}^*(f)$ is a rank (*n*-1) fibre bundle over $(c_i, c_{i+1}) \hat{I} R(f)$ and we say that $\boldsymbol{b}_M^*(f) = \underset{i,j}{\oplus} \boldsymbol{b}_{M_i^j}^*(f)$ is the fibre bundle local structure induced by *f* on *M*.

Pick now a connected component $V_c^i$ ($c \in R$, $i \in I(c)$) such that $f^{-1}(c)$ is non empty and $f$ is submersive along it ($df|_{V_c^i} \neq 0$). Obviously $V_c^i$ is an analytic submanifold on $M$. Let $G$ be an $f$-partitioned open subset containing $V_c^i$ where $f$ is submersive (it always exists by continuity). In certain subset $V \subset G$ ($V \cap V_c^i \neq \emptyset$) the equilibrium condition implies that $\|\nabla f\|^2 = w(f)$ and $\Delta f = y(f)$ ($w$, $y$ analytic functions of its argument, see the proof of theorem 3.6). In fact as the partition $b_\Gamma(f)$ is locally trivial it is easy to see that the functional dependences $\|\nabla f\|^2 = w(f)$ and $\Delta f = y(f)$ hold in the whole $G$. We have then that the mean curvature $H$ (see equation (2.9)) has the following expression

$$H = div\left(\frac{\nabla f}{\|\nabla f\|}\right) = \frac{-1}{\|\nabla f\|^2}\nabla(\|\nabla f\|)\nabla f + \frac{\Delta f}{\|\nabla f\|} = \frac{-w'(f)}{2\sqrt{w(f)}} + \frac{y(f)}{\sqrt{w(f)}} \quad . \tag{5.1}$$

Equation (5.1) implies that the mean curvature is constant on all regular $V_c^i$.

Now we prove the geodesical parallelism between two near connected components of the fibres of $f$ in $G$ ($V_{c_1}^{i_1}$ and $V_{c_2}^{i_2}$ with $c_1$ close to $c_2$). By taking the covariant derivative along $X \in T^w(M)$ we have the following equalities

$$D_X g(\nabla f, \nabla f) = D_X w(f) = w'(f)(\nabla f)^j X_j \quad . \tag{5.2}$$

$$D_X g(\nabla f, \nabla f) = 2g(D_X \nabla f, \nabla f) \stackrel{[12]}{=} 2(\nabla f)_{;k}^j (\nabla f)^k X_j \quad . \tag{5.3}$$

Identifying (5.2) and (5.3) we obtain that

$$g(D_{\nabla f}\nabla f, X) = g(\frac{w'(f)}{2}\nabla f, X) \quad \Rightarrow \quad D_{\nabla f}\nabla f = \frac{w'(f)}{2}\nabla f \tag{5.4}$$

which is the condition on the integral curves of $\nabla f$ to be geodesics in $G$ (although in a parameter $l$ which is not the arc length). Let us see that the variation of $l$ as we move from the hypersurface $V_{c_1}^{i_1}$ to $V_{c_2}^{i_2}$ is independent of the integral curve of $\nabla f$ that we choose to do so. Indeed

$$\frac{df}{dl} = \|\nabla f\|^2 = w(f) \quad \Rightarrow \quad \int_{c_1}^{c_2} \frac{df}{w(f)} = \Delta l \tag{5.5}$$

which depends exclusively on the $i_1$ connected component of $f^{-1}(c_1)$ and the $i_2$ connected component of $f^{-1}(c_2)$ and not on the path chosen. Since the arc length is related to the parameter $l$ by the expression $ds = \sqrt{w(f(l))}dl$ we have that the distance between $V_{c_1}^{i_1}$ and $V_{c_2}^{i_2}$ along the integral curves of $\nabla f$ only depends on the initial and the final hypersurfaces. Therefore due to the geodesical character of the integral curves of $\nabla f$ we have that $V_{c_1}^{i_1}$ and $V_{c_2}^{i_2}$ are, by definition [7], geodesically parallel. □

## 3-dimensional results independent of the topology.

Theorem 5.1 gives a characterization of an equilibrium partition on a general Riemannian manifold. Now we are interested in giving further properties of the regular hypersurfaces $S \in \boldsymbol{b}_M(f)$ in more specific spaces and eventually in obtaining a complete clasification of $\boldsymbol{b}_M(f)$ for any equilibrium function $f$ on $M$. The more we want to characterize an equilibrium partition $\boldsymbol{b}_M(f)$ the more we have to restrict the topology and geometry of the base space $M$.

**Proposition 5.2.** Let $S$ be a codimension 1 submanifold of $M$ (dim($M$) = 3) such that verifies the following

(i) $R$ is constant on $S$.
(ii) $R'$ is constant on $S$.
(iii) $R_{ab} n^a n^b$ is constant on $S$.

Then $S$ has constant Gauss curvature.

*Proof.* Let $u, v$ be two orthonormal vectors tangent to $S$ at the point $p \hat{I} S$. The sectional curvature $K$ of $M$ restricted to $S$ can be computed by using equation (2.8) and the expression of the Riemann tensor of a 3-dimensional Riemannian manifold [7].

$$K = R_{abcd} u^a v^b u^c v^d = \left( R_{ac} g_{bd} - R_{ad} g_{bc} + R_{bd} g_{ac} - R_{bc} g_{ad} + \frac{R}{2} g_{ad} g_{bc} - \frac{R}{2} g_{ac} g_{bd} \right) u^a v^b u^c v^d =$$

$$= R_{ab} \left( u^a u^b + v^a v^b \right) - \frac{R}{2} .$$

The expression $u^a u^b + v^a v^b$ is a projection tensor onto $S$ and hence $u^a u^b + v^c v^d = \boldsymbol{b}^{ab}$. Taking this fact into account and considering the expression of $\boldsymbol{b}^{ab}$ (recall section 2) we get the following equation for $K$.

$$K = \frac{R}{2} - R_{ab} n^a n^b \qquad . \tag{5.6}$$

By assumptions (i) and (iii) we have that $K$ is constant on $S$.
Now we show that $K'$ is also constant on $S$. Indeed by using the expression of the Riemann tensor of a 2-dimensional Riemannian manifold [7] we get

$$K' = R'_{abcd} u^a v^b u^c v^d = \frac{R'}{2} \tag{5.7}$$

and therefore by assumption (ii) $K'$ is constant on $S$. As a consequence of Gauss theorem (equation (2.7)) $\overline{K}$ is constant on $S$ and we are done. □

*Proof of theorem 3.7 for flat and conformally flat manifolds.* Now there remains to prove that the Gauss curvature $\overline{K}$ is constant on each fibre of $\boldsymbol{b}_M^*(f)$. We work in a specific local coordinate system $(x^1, x^2, x^3)$ where the metric tensor of the conformally flat space has the canonical form $g_{ab} = e^{2\boldsymbol{f}} \boldsymbol{d}_{ab}$. Furthermore we suppose that the equilibrium function $f$ on $M$ and the function $\boldsymbol{f}$ agree fibrewise. Note that a flat space is obtained from a conformally flat space by taking $\boldsymbol{f}=0$, in which case the coupling assumption is always verified. The Ricci tensor and the scalar curvature of $(M,g)$ in terms of $\boldsymbol{f}$ are given by [7]

$$R_{ab} = \boldsymbol{f}_{,ab} - \boldsymbol{f}_{,a}\boldsymbol{f}_{,b} + \boldsymbol{d}_{ab}\left(\Delta_E \boldsymbol{f} + \|\nabla_E \boldsymbol{f}\|_E^2\right) \qquad (5.8)$$

$$R = e^{-2\boldsymbol{f}}\left(4\Delta_E \boldsymbol{f} + 2\|\nabla_E \boldsymbol{f}\|_E^2\right) \qquad (5.9)$$

where the subscript $E$ means that the corresponding operator has the Euclidean form in the local coordinate system $(x^i)$. We work in an $f$-partitioned open, connected subset $\Gamma \subset M_i^j$ (where $f$ is submersive) where we can express locally that $\boldsymbol{f} = \Phi(f)$ (note that $G$ always exists due to the local triviality of the partition of $M_i^j$). The globalization of the local results obtained in this way is achieved by employing lemma 2.9. For the sake of simplicity we fix our attention in a surface $S \subset G$. It is immediate to check that $\boldsymbol{f}$ is an equilibrium function and that

$$\|\nabla \boldsymbol{f}\|^2 = e^{-2\boldsymbol{f}} \|\nabla_E \boldsymbol{f}\|_E^2 \qquad (5.10)$$

$$\Delta \boldsymbol{f} = e^{-2\boldsymbol{f}}\left(\|\nabla_E \boldsymbol{f}\|_E^2 + \Delta_E \boldsymbol{f}\right) \qquad (5.11)$$

From equations (5.9), (5.10) and (5.11) it is evident that $R$ is constant on $S$. Let us now prove that $R'$ is also constant. Note by looking at equation (2.6) that if $H_{ab}H^{ab}$ and $R_{ab}n^a n^b$ are both constant on $S$ then $R'$ is as well.

$$H_{ab}H^{ab} = \frac{1}{4}\left(L_{\tilde{n}}\boldsymbol{b}_{ab}\right)\left(L_{\tilde{n}}\boldsymbol{b}^{ab}\right) = \frac{1}{4}[3\left(L_{\tilde{n}}e^{2\boldsymbol{f}}\right)^2 - \left(L_{\tilde{n}}e^{2\boldsymbol{f}}\right)\left(L_{\tilde{n}}\tilde{n}^a\right)\tilde{n}_a - \left(L_{\tilde{n}}e^{2\boldsymbol{f}}\right)\left(L_{\tilde{n}}\tilde{n}^b\right)\tilde{n}_b -$$
$$- \left(L_{\tilde{n}}e^{2\boldsymbol{f}}\right)\left(L_{\tilde{n}}\tilde{n}_a\right)\tilde{n}^a + \left(L_{\tilde{n}}\tilde{n}_a\right)\left(L_{\tilde{n}}\tilde{n}^a\right) + \left(L_{\tilde{n}}\tilde{n}_a\right)\tilde{n}^a\left(L_{\tilde{n}}\tilde{n}_b\right)\tilde{n}^b - \left(L_{\tilde{n}}e^{2\boldsymbol{f}}\right)\left(L_{\tilde{n}}\tilde{n}_b\right)\tilde{n}_a \boldsymbol{d}^{ab} +$$
$$+ \left(L_{\tilde{n}}\tilde{n}^a\right)\tilde{n}_a\left(L_{\tilde{n}}\tilde{n}_b\right)\tilde{n}^b + \left(L_{\tilde{n}}\tilde{n}_b\right)\left(L_{\tilde{n}}\tilde{n}^b\right)].$$

As $\left(L_{\tilde{n}}\tilde{n}^a\right)\tilde{n}_a = \frac{1}{2}L_{\tilde{n}}\left(\tilde{n}^a\tilde{n}_a\right) \stackrel{\tilde{n}^a\tilde{n}_a=1}{=} 0$ the above expression simplifies to

$$H_{ab}H^{ab} = \frac{1}{4}\left[3\left(L_{\tilde{n}}e^{2\boldsymbol{f}}\right)^2 + 2\left(L_{\tilde{n}}\tilde{n}_a\right)\left(L_{\tilde{n}}\tilde{n}^a\right)\right] \qquad (5.12)$$

It is easy to check that the first term in (5.12) is constant on $S$ using the expression $\tilde{n}^a = \dfrac{(\nabla f)^a}{\|\nabla f\|}$ for the normal vector field. For the second term notice the following computation

$$\left(L_{\tilde{n}}\tilde{n}_{a}\right)\left(L_{\tilde{n}}\tilde{n}^{a}\right) = \|\nabla f\|^2 \left(L_{\tilde{n}} \frac{1}{\|\nabla f\|}\right)^2 + \frac{1}{\|\nabla f\|}\left(L_{\tilde{n}} \frac{1}{\|\nabla f\|}\right)\left(L_{\tilde{n}} \|\nabla f\|^2\right) + \frac{1}{\|\nabla f\|^2}\left(L_{\tilde{n}}(\nabla f)^a\right)\left(L_{\tilde{n}}(\nabla f)_a\right).$$

The first and the second terms are trivially constant on *S*. The third one requires more computations.

$$\left(L_{\tilde{n}}(\nabla f)^a\right)\left(L_{\tilde{n}}(\nabla f)_a\right) = \frac{1}{\|\nabla f\|^2}(\nabla f)^b (\nabla f)^c \frac{\partial(\nabla f)_a}{\partial x^b}\frac{\partial(\nabla f)^a}{\partial x^c} = \frac{1}{\|\nabla f\|^2}e^{-4f}\frac{\partial f}{\partial x^b}\frac{\partial f}{\partial x^c}\frac{\partial^2 f}{\partial x^a \partial x^b}$$

$$\cdot\left[\frac{\partial e^{-2f}}{\partial x^c}\frac{\partial f}{\partial x^a} + e^{-2f}\frac{\partial^2 f}{\partial x^c \partial x^a}\right].$$

On the other hand $\frac{\partial e^{-2f}}{\partial x^c} = -2e^{-2f}\Phi'(f)\frac{\partial f}{\partial x^c}$ and $\frac{\partial^2 f}{\partial x^a \partial x^b}\frac{\partial f}{\partial x^a} = \frac{1}{2}\frac{\partial}{\partial x^b}\left(\|\nabla_E f\|_E^2\right)$. Substituting these expressions into the above equation we get a formula for $\left(L_{\tilde{n}}(\nabla f)^a\right)\left(L_{\tilde{n}}(\nabla f)_a\right)$, which after some computations is clearly constant on *S* and hence $H_{ab}H^{ab}$ is also constant.

The computations to prove that $R_{ab}n^a n^b$ is constant on *S* are similar. For the Ricci tensor we use the expression (5.8):

$$R_{ab}n^a n^b = \frac{e^{-4f}}{\|\nabla f\|^2}\left[\frac{\partial^2 \boldsymbol{f}}{\partial x^a \partial x^b} - \frac{\partial \boldsymbol{f}}{\partial x^a}\frac{\partial \boldsymbol{f}}{\partial x^b} + \boldsymbol{d}_{ab}\left(\Delta_E \boldsymbol{f} + \|\nabla_E \boldsymbol{f}\|_E^2\right)\right]\frac{\partial f}{\partial x^a}\frac{\partial f}{\partial x^b} \qquad (5.13)$$

Some computations in (5.13) yield immediately to prove that $R_{ab}n^a n^b$ is constant on *S* and by (2.6) *R'* is also constant as we wanted to prove. Note now that proposition 5.2 can be applied and hence Gauss curvature is also constant on *S*. The arguments above extend to every fibre of $\boldsymbol{b}_M^*(f)$. □

*Remark 1.* Note that the above proof did not make use of the topology of *M*. This is due to the local character of our computations.

*Remark 2.* If we have an equilibrium function on a conformally flat manifold *M*, but the coupling assumption is not satisfied, the theorem does not apply. Therefore it remains an open problem to give further geometrical characterization (must the Gauss curvature be constant on each regular surface?). Note that if *M* is flat the coupling condition does not play a role and hence the theorem 3.7 holds for any flat 3-manifold.

*Proof of theorem 3.7 for locally symmetric manifolds.* As in the first part of the proof we only have to show that the Gauss curvature is constant on each fibre of $\boldsymbol{b}_M^*(f)$. For this kind of spaces the coupling condition that we impose is that S must have parallel second fundamental form [13] ($H_{ab|c} = 0$ on *S*). Recall that *M* is called locally symmetric [14] if $R_{abcd;m} = 0$. Condition (i) of proposition 5.2 is immediately verified, *R* is constant on *S*. Let us now show that condition (iii) is also satisfied. The induced metric tensor on $S$, $\boldsymbol{b}^{ab}$, is covariantly constant with respect to the induced covariant derivative

$$0 = \boldsymbol{b}^{ab}_{\|e} = \left(g^{ij} - \tilde{n}^i\tilde{n}^j\right)_{;m} \boldsymbol{b}^a_i \boldsymbol{b}^b_j \boldsymbol{b}^m_e = -\left(\tilde{n}^i\tilde{n}^j\right)_{;m} \boldsymbol{b}^a_i \boldsymbol{b}^b_j \boldsymbol{b}^m_e = -\left(n^a n^b\right)_{\|e}. \qquad (5.14)$$

If now we take in $R_{ab}n^a n^b$ the induced covariant derivative and use expression (5.14) we get $\left(R_{ab}n^a n^b\right)_{\|e} = R_{ab}\left(n^a n^b\right)_{\|e} = 0$ which means that $R_{ab}n^a n^b$ is constant on $S$. Finally condition (ii) is an easy consequence of $\left(H_{ab}H^{ab}\right)_{\|e} = 0$ due to the fact that the second fundamental form of $S$ is parallel and by (2.6) it is immediate that $R'$ must be constant on $S$. Again applying proposition 5.2 we obtain that the Gauss curvature $\bar{K}$ and therefore the principal curvatures are constant on $S$. □

*Remark 3.* The geometrical restriction on the equilibrium partitions of locally symmetric 3-manifolds obtained in the previous proof has been obtained by considering certain coupling between the metric tensor and the fibres of $\boldsymbol{b}^*_M(f)$. The same was necessary for conformally flat spaces. These facts suggest that the strongest geometrical restrictions on equilibrium partitions of manifolds appear when the metric $g$ of $M$ is coupled in some way with the regular surfaces of $\boldsymbol{b}_M(f)$.

### Results for n-manifolds diffeomorphic to $R^n$.

In this final subsection we obtain the most restrictive classifications of an equilibrium partition on a manifold. The restrictions are not only local as in the previous theorems but global and valid for every connected component of $f^{-1}(c)$ ($c\hat{I}f(M)$) (regular or critical). The base manifold $M$ is $n$-dimensional in this case but we have to restrict its topology, $M$ must be diffeomorphic to $R^n$.

*Proof of theorem 3.8 for flat manifolds.* Now we want to prove that the principal curvatures are constant on each fibre. As usual we consider an $f$-partitioned open connected subset $\boldsymbol{G}$ in some $M^j_i$. Pick two close hypersurfaces $S$ and $S'$ of $\boldsymbol{b}_M(f)$ in $\boldsymbol{G}$ separated by a geodesical distance $r$. The principal curvatures $k_i$ and $k_i'$ at corresponding points $p\hat{I}S$ and $p'\hat{I}S'$ ($p$ and $p'$ are joint by a straight line orthogonal to $S$ and $S'$) are related by the following formula [15]

$$k'_i = \frac{k_i}{1+rk_i} \qquad . \qquad (5.15)$$

From the definition of the mean curvature in terms of the principal curvatures $H = \sum_{i=1}^{n-1} k_i$ and after some computations it is easy to obtain this equation

$$H' = \frac{1}{\prod_{i=1}^{n-1}(1+rk_i)}\left[H + \left(\sum_{i\neq j} 2k_i k_j\right)r + \left(\sum_{i\neq j\neq l} 3k_i k_j k_l\right)r^2 + \ldots + (n-1)\left(\prod_{i=1}^{n-1} k_i\right)r^{n-2}\right]. \qquad (5.16)$$

As $H'$ is constant on each hypersurface of $S' \in b_M(f)$ in $G$ we have that $H' = H'(r)$. Let us introduce the following coordinate system in $G$: $(x^1,...,x^n) \to (r, r'_1,..., r'_{n-1})$ where $r$ is the distance function [16] to a fixed hypersurface $S$ (which is analytic if $G$ is small enough) and the other coordinates are introduced to guarantee that

$$rk \begin{pmatrix} \frac{\partial r}{\partial x^1} & ... & \frac{\partial r}{\partial x^n} \\ \vdots & ... & \vdots \\ \frac{\partial r'_{n-1}}{\partial x^1} & ... & \frac{\partial r'_{n-1}}{\partial x^n} \end{pmatrix} = n \quad \text{in } \Gamma. \tag{5.17}$$

Condition (5.17) may not be satisfied in the whole $G$ but there always exists an open subset $U \subset \Gamma$ for which the coordinates are well defined. In this local coordinate system the principal curvatures $k_i$ are functions of $r'_1,..., r'_{n-1}$. Now we take derivatives with respect to $r$ on both sides of equation (5.16) and evaluate at $r=0$. After some computations it is easy to obtain that the functions

$$\sum_{i=1}^{n-1} k_i, \sum_{i=1}^{n-1} k_i^2, ..., \sum_{i=1}^{n-1} k_i^{n-1} \tag{5.18}$$

must be constant on $S \cap U$ and by analyticity constant on the whole $S$. As we have $n-1$ (independent) conditions for $n-1$ functions we must conclude that the principal curvatures of $S$ do not depend on $r'_i$ and therefore are constant real numbers. We extend this property to all the fibres of $M_i^j$ by using equation (5.15). The conclusion is that the principal curvatures of all the fibres of $b_M^*(f)$ are constant on each fibre.

The complete, connected, codimension 1 submanifolds in $(R^n, \text{flat})$ with constant principal curvatures are classified [17]. They can only be (globally) isometric to $R^{n-1}$, $S^{n-1}$ and $S^{n-1-k} \times R^k$ ($1 \pounds k \pounds n-2$) with their respective canonical metrics. Due to the fibre bundle local structure of $b_M(f)$ and the analiticity (see lemma 2.9) it is easy to see that all the components (regular or critical) of the partition induced by $f$ must be (globally) isometric to a standard cylinder. ☐

*Remark 4.* The theorem in [17] which allowed this clasification does not apply in general if the topology of the base space $(M, \text{flat})$ is not $R^n$. This fact manifests that an almost-trivial partition is not necessarily geometrically trivial and that the equivalence between both concepts depends on the topology of the base space.

*Proof of theorem 3.8 for conformally flat manifolds.* As $f$ is an equilibrium function it is immediate to prove that $f, \|\nabla_E f\|_E^2$ and $\Delta_E f$ agree fibrewise. Taking this fact into account let us construct the following isometric embedding. The map $\aleph: b_M(f) \to b_{R^n}(f)$ assigns to each connected component of the fibres of $f$ in $(M,g)$ the same connected component of the fibres of $f$ in $(R^n, d)$. $\tilde{A}$ is a diffeomorphism and furthermore an isometric embedding. This is a consequence of that the induced metrics on $S \hat{I} b_M(f)$ and $\aleph(S) \hat{I} b_{R^n}(f)$ are the same except for a constant positive global factor. As $f, \|\nabla_E f\|_E^2$ and $\Delta_E f$ agree fibrewise the partition on $(R^n, \boldsymbol{d})$ is geometrically trivial and hence $\boldsymbol{b}_M(f)$ is as well. $\square$

*Remark 5.* The technique used above does not only work for $M$ diffeomorphic to $R^n$ but it is immediate to prove that $\boldsymbol{b}_M(f)$, where $M$ is a conformally flat $n$-manifold and the coupling condition holds, can be (globally) isometrically embedded into $(M, \boldsymbol{d})$.

*Remark 6 (constant curvature spaces).* The three canonical constant curvature spaces are $R^n$, $H^n$ and $S^n$. For the first one theorem 3.8 gives a complete classification of the equilibrium partitions of it. By using the generalized Alexandrov's theorem [18] we can obtain an almost complete classification for the other two spaces. As a consequence of this theorem it is immediate to prove that if $M=H^n$ and one of the components of the equilibrium partition on $M$ is compact then all the components of $\boldsymbol{b}_M(f)$ must be (globally) isometric to $bS^{n-1}$ (note that we do not need to impose the coupling condition). We cannot say whether or not $\boldsymbol{b}_M(f)$ is geometrically trivial when all the components of the equilibrium partition are not compact. The same happens if $M=S^n$ but in this case the components of $\boldsymbol{b}_M(f)$ are necessarily compact and therefore the classification is total (without coupling assumption): all the components of $\boldsymbol{b}_M(f)$ must be (globally) isometric to $bS^{n-1}$.

*Remark 7.* All the results obtained in this section manifest that the topological and geometrical properties of an equilibrium partition depend on the base manifold $(M,g)$ on which it is defined. If we want the equilibrium partition to be almost-trivial, geometrically trivial or another interesting definition we have to restrict the geometry and topology of $(M,g)$.

*Final remark.* As an open problem it would be interesting to know whether or not the coupling conditions used in the proof of the classification theorems for conformally flat and locally symmetric spaces are necessary in order that equilibrium functions exist.

## 6. Equilibrium functions and isometries.

Apart from the classificatory results obtained in the preceding section the equilibrium partitions of manifolds have other interesting properties that show the deep geometrical meaning of this kind of partitions and their strong relation with the base space $(M,g)$.

In the Euclidean space $R^n$ the group of isometries $G$ is generated by the transformations in the base $\mathcal{B}$ of the Lie algebra $\mathcal{G}$ (in cartesian coordinates $(x^1,...,x^n)$)

$$\mathcal{B} = \left\{\partial_{x^1},...,\partial_{x^n}, x^1\partial_{x^2} - x^2\partial_{x^1},..., x^1\partial_{x^n} - x^n\partial_{x^1}, x^2\partial_{x^3} - x^3\partial_{x^2},..., x^{n-1}\partial_{x^n} - x^n\partial_{x^{n-1}}\right\}. \quad (6.1)$$

Let $\mathcal{G}_{pq}$ be a Lie subalgebra of $\mathcal{G}$ with $p$ generators $X_1,...,X_p$ and $rk(X_1,...,X_p) = q$ ($p \geq q$). We are interested in the ($n$-1)-rank subgroups of $G$. We can generate these subgroups by taking $p \geq n$-1 vector fields $X_1,...,X_p$ constant coefficient linear combinations of elements in $\mathcal{B}$ satisfying the following

$$\left.\begin{array}{l} \left[X_i, X_j\right] = c_{ijk} X_k \\ c_{ijk} \in R, \ 1 \leq i,j,k \leq p \\ rk(X_1,...,X_p) = n-1 \text{ up to a null measure set} \end{array}\right\} \quad (6.2)$$

**Definition 6.1.** We say that a partition $\boldsymbol{b}_{R^n}(f)$ is induced by a ($n$-1)-rank subgroup of $G$ if the corresponding generators $X_1,...,X_p$ of this subgroup (satisfying (6.2)) verify that $X_i(f) = 0$ ($i = 1,..., p$).

Each one of the regular connected components of $f$ is an analytic codimension 1 submanifold of $R^n$ [19] with the property of being a transitivity hypersurface of the ($n$-1)-rank subgroup.

Let us now prove that all these transitivity hypersurfaces are in fact equilibrium submanifolds.

**Theorem 6.2.** The partitions induced by any ($n$-1)-rank subgroup of the Euclidean group of isometries $G$ are equilibrium partitions of $(R^n, \boldsymbol{d})$.

*Proof.* For simplicity we will work in $R^3$ although the results to be obtained are valid in $R^n$. Here $x^1 = x$, $x^2 = y$ and $x^3 = z$. Suppose an analytic function $f$ on $R^3$ and a 2-rank subgroup of $G$ of generators $X_1,...,X_p$. Assume that the partition $\boldsymbol{b}_{R^3}(f)$ is induced by this subgroup and hence $X_1(f) = 0,..., X_p(f) = 0$. It is a standard result [14] that the isometries of a Riemannian manifold commute with the Laplace-Beltrami operator, in particular, the elements of $\mathcal{B} = \left\{\partial_x, \partial_y, \partial_z, x\partial_y - y\partial_x, x\partial_z - z\partial_x, y\partial_z - z\partial_y\right\}$ commute with $\Delta = \dfrac{\partial^2}{\partial x^2} + \dfrac{\partial^2}{\partial y^2} + \dfrac{\partial^2}{\partial z^2}$. Hence we have

$$X_i(\Delta f) = \Delta(X_i(f)) = 0, \; i = 1,\ldots,p \qquad (6.3)$$

and therefore $\Delta f$ and $f$ agree fibrewise.

Now let us check that $X_i(\nabla f \cdot \nabla) = (\nabla f \cdot \nabla) X_i$. It is enough to verify this condition for $\partial_z$ and $x\partial_y - y\partial_x$, for the rest of vector fields in $\mathcal{B}$ and the linear combinations of them is completely analogous. If $X = \partial_z$ we get

$$\partial_z(f_{,x}\partial_x + f_{,y}\partial_y + f_{,z}\partial_z) = f_{,x}\partial_{xz} + f_{,y}\partial_{yz} \text{ when } \partial_z f = 0 \qquad (6.4)$$

$$\left(f_{,x}\partial_x + f_{,y}\partial_y + f_{,z}\partial_z\right)\partial_z = f_{,x}\partial_{zx} + f_{,y}\partial_{zy} \text{ when } \partial_z f = 0 \quad . \qquad (6.5)$$

So (6.4) and (6.5) are the same operator. If $X = x\partial_y - y\partial_x$ we have the following equations

$$(x\partial_y - y\partial_x)\left(f_{,x}\partial_x + f_{,y}\partial_y + f_{,z}\partial_z\right) = x\left(f_{,x}\partial_{xy} + f_{,y}\partial_{yy} + f_{,z}\partial_{zy}\right) -$$
$$- y\left(f_{,x}\partial_{xx} + f_{,y}\partial_{yx} + f_{,z}\partial_{zx}\right) + \left(xf_{,xy} - yf_{,xx}\right)\partial_x + \left(xf_{,yy} - yf_{,yx}\right)\partial_y + \qquad (6.6)$$
$$+ \left(xf_{,zy} - yf_{,zx}\right)\partial_z$$

and

$$\left(f_{,x}\partial_x + f_{,y}\partial_y + f_{,z}\partial_z\right)(x\partial_y - y\partial_x) = x\left(f_{,x}\partial_{yx} + f_{,y}\partial_{yy} + f_{,z}\partial_{yz}\right) -$$
$$- y\left(f_{,x}\partial_{xx} + f_{,y}\partial_{xy} + f_{,z}\partial_{xz}\right) + f_{,x}\partial_y - f_{,y}\partial_x \qquad (6.7)$$

Taking into account that $X(f) = xf_{,y} - yf_{,x} = 0$ and taking derivatives of it with respect to $x, y$ and $z$ we have

$$\left.\begin{array}{r}xf_{,xy} - yf_{,xx} = -f_{,y} \\ xf_{,yy} - yf_{,yx} = f_{,x} \\ xf_{,zy} - yf_{,zx} = 0\end{array}\right\} \quad . \qquad (6.8)$$

Substituting (6.8) into (6.6) it is immediate to get the same expression that in (6.7) and hence the commutation is proved. Therefore we can write in general for any generators $X_i$ that

$$X_i(\|\nabla f\|^2) = (\nabla f \cdot \nabla) X_i(f) = 0 \quad . \qquad (6.9)$$

On account of (6.9) we have that $\|\nabla f\|^2$ and $f$ agree fibrewise. As $\Delta f$ and $f$ agree fibrewise as well we have that $\boldsymbol{b}_{R^3}(f)$ is an equilibrium partition. □

*Remark 1.* An immediate consequence of theorem 6.2 is that all the partitions induced by $(n-1)$-rank subgroups of $G$ must be geometrically trivial (theorem 3.8). Note also that we have not used the classification theorem 3.8 in the proof of theorem 6.2. By using the classification theorem 3.8 one can obtain the implication of theorem 6.2 in the opposite direction: all the equilibrium partitions of ($R^n$, **d**) are induced by ($n$-1)-rank subgroups of isometries.

*Remark 2*. The proof of the corollary appearing in the above remark can be done in a different but much more involved manner. Take the vector fields $X_i$ as general linear combinations of the elements of $\mathcal{B}$ and study the condition $X_i(f) = 0$ by solving the corresponding PDE system $\{X_i \nabla f = 0\}_{i=1,...,n}$ case by case (for the different possible choices of $\{X_i\}_{i=1,...,p}$ satisfying (6.2)). As the reader immediately can note this approach is more involved than the proof of theorem 6.2.

*Final remark*. Theorem 6.2 is specific for the Euclidean space $R^n$ but it is natural to try to extend it to more general Riemannian manifolds. If ($M$,$g$) has a group $G$ of isometries with ($n$-1)-rank subgroups, is it true that all the partitions induced by these subgroups are equilibrium partitions and conversely?. If ($M$,$g$) does not possess isometries or its group of isometries does not possess at least ($n$-1) generators satisfying (6.2) the question is whether equilibrium partitions exist on such spaces.

# 7. Physical applications: static self-gravitating fluids.

In this section we review Newtonian and relativistic fluids comparing the implications of this paper with the classical techniques and results obtained by other authors.

## Self-gravitating Newtonian fluids.

The gravitatory potential function $V: R^3 \to R$ and the density and pressure functions $r, p: \Omega \to R$ are to satisfy equations of problem (P2)

$$\left. \begin{array}{l} \Delta V = r \quad \text{in } \Omega \\ \nabla p + r \nabla V = 0 \quad \text{in } \Omega \\ V = c \ (c \in R), \nabla V \neq 0 \text{ and } V \in C_t^2 \quad \text{on } \partial\Omega \\ \Delta V = 0 \quad \text{in } R^3 - \overline{\Omega} \end{array} \right\} \qquad (7.1\text{-}7.4)$$

where the domain $\Omega$ is a connected open set which is regular enough (for instance requiring $\partial\Omega$ to be a smooth submanifold) and $r$ and $p$ are smooth functions on $\Omega$ and constant on $\partial\Omega$. Should $\Omega$ is bounded we usually require $V$ to approach zero as we tend to infinity.

In what follows and for adapting the notation to the terminology employed in the fluid mechanics literature, we will say that $\Omega$ is a perfect sphere when $\partial\Omega$ is (globally) isometric to $bS^2$ ($bS^{n-1}$), a perfect cylinder when $\partial\Omega$ is (globally) isometric to $bS^1 \times R$ ($bS^{n-1-k} \times R^k, k = 1,...,n-2$) and a region bounded by parallel planes when $\partial\Omega$ has two parallel (geodesically parallel) connected components (globally) isometric to $R^2$ ($R^{n-1}$).

When $\Omega$ is a bounded domain Lichtenstein proved a theorem for rotating self-gravitating fluids which in the static case implies the existence of infinite planes of mirror symmetry [20]. As a consequence of this fact $\Omega$ must be a perfect sphere. The most important physical assumptions imposed by Lichtenstein are that $V$ must tend to zero at infinity and $r$ must be a non-negative function on $\Omega$. The proof has an analytical nature and is only valid for bounded domains.

More recently Lindblom [21] gave a different proof of the existence of infinite planes of mirror symmetry. It is again of analytical nature and the hypotheses imposed are:

(a) The existence of a state equation $r = r(p)$ such that $\dfrac{dr}{dp} \geq 0$.

(b) $r \geq 0$, $p \geq 0$ and the asymptotic condition on $V$, $\lim_{\infty} V = 0$.

Lindblom's proof confirms that when $\Omega$ is bounded and the (a) and (b) conditions hold then the only solutions to (P2) are functions with spherical symmetry. His proof fails for unbounded regions $\Omega$.

The techniques developed in this paper are advantageous with respect to the classical approaches. Firstly note that the equations of (P2) are a particular case of the equations of (P1) and that the regularity conditions in (P1) are fulfilled by the physical assumptions in (P2), supposing the analiticity of $r$ and $p$ on $\Omega$. Let us summary now the main contributions of this paper to the study of the Newtonian free-boundary problem:

(i) We study both bounded and unbounded domains $\Omega$ in the same framework. The classical approaches (specifically Lichtenstein's and Lindblom's proofs) are only valid for bounded domains.

(ii) We solve from a geometrical point of view, and without imposing any physical assumption (as (a) or (b)), the problem of classifying the equilibrium shapes of self-gravitating Newtonian fluids in the Euclidean $R^n$. The classical proofs are of analytical nature, valid in $R^3$ and impose physical assumptions.

(iii) The classification extends to more general Riemannian spaces, see theorems 3.7, 3.8 and 5.1. These cases correspond to the model of a Newtonian fluid on a Riemannian manifold. It is difficult that we be able to obtain these results through the classical approaches. On the other hand, the classical approaches (Lichtenstein, Lindblom …) are adapted to prove the physical intuition of spherical symmetry and hence they do not say anything of the spaces on which the equilibrium partition (the partition induced by the potential function $V$) is not geometrically trivial. On the contrary our approach is not adapted to specific partitions but it is valid and conclusive whether $\Omega$ has to be a perfect sphere or not. The technique developed in this paper of substituting a system of PDE by the equilibrium partition condition (theorem 3.6) lays down the foundations of a general theory of equilibrium shapes in general geometries and topologies of the base manifold.

## Self-gravitating relativistic fluids.

Without loss of generality assume that the space-time manifold is diffeomorphic to $M \mathrm{x} R$. The space-time metric can be written in local coordinates like $ds^2 = -V^2 dt^2 + g_{ab} dx^a dx^b$, where $V$ is the potential function on ($M$, $g_{ab}$) and satisfies the following equations which we set up as problem (P3)

$$\left. \begin{array}{l} \Delta V = 4\pi V(r+3p) \quad \text{in } \Omega \\ V\nabla p + (r+p)\nabla V = 0 \quad \text{in } \Omega \\ V = c \ (c \in R), \nabla V \neq 0 \text{ and } V \in C_t^2 \quad \text{on } \partial\Omega \\ \Delta V = 0 \quad \text{in } M - \overline{\Omega} \end{array} \right\} . \qquad (7.5\text{-}7.8)$$

Analogously to the Newtonian case the domain $\Omega$ is a connected open set which is regular enough (for instance requiring $\partial\Omega$ to be a smooth submanifold) and $r$ and $p$ are smooth functions on $\Omega$ and constant on $\partial\Omega$.

When $\Omega$ is bounded (and $M$ is non-compact), a couple of standard asymptotic conditions are usually imposed

$$\left. \begin{array}{l} V = 1 - \dfrac{m}{r} + o(r^{-2}) \\[2mm] g_{ab} = \left(1 + \dfrac{2m}{r}\right) d_{ab} + o(r^{-2}) \end{array} \right\} \quad (7.9,\ 7.10)$$

$m$ standing for the mass of the fluid-composed star and $r$ for the asymptotic radial coordinate. The metric tensor $g_{ab}$ is not free but it is coupled with the matter via the following formula

$$R_{ab} = V^{-1} V_{;ab} + 4\boldsymbol{p}(\boldsymbol{r} - p) g_{ab} \ . \tag{7.11}$$

For the sake of simplicity we impose that $\boldsymbol{r}$ and $p$ are analytic on $\Omega$. The metric tensor $g_{ab}$ is analytic on $\Omega$ and $M - \bar{\Omega}$ as Muller zum Hagen has proved [22]. These assumptions automatically imply that $V$ is also analytic on the outer and inner regions (see lemma 3.1). The metric tensor cannot be analytic across the free-boundary $\partial\Omega$, as one can see immediately by computing the scalar curvature from (7.11), but it must be $C_t^2$ on the boundary on account of the junction conditions of Synge [23].

We then ask the same question as in the Newtonian case, say, are there conditions on $\Omega$ or on the symmetries of $\boldsymbol{r}, p, V$ and $g_{ab}$ so that (P3) has a solution? What are the possible equilibrium shapes of a relativistic self-gravitating fluid?

Note the reader that this case is much more difficult than the Newtonian case since the metric tensor is coupled with the matter through equation (7.11) and is not free. The metric tensor is itself an unknown of the problem.

The best result so far to our knowledge is the one obtained by Beig and Simon [24]. Their original idea was later improved by Lindblom and Masood-ul-Alam [25] by removing certain technical condition from [24]. They have given a proof of the spherical symmetry of the solutions to (P3) when this series of hypotheses hold

(a) Existence of a state equation $\boldsymbol{r} = \boldsymbol{r}(p)$ such that $\dfrac{d\boldsymbol{r}}{dp} \geq 0$.

(b) $\boldsymbol{r} \geq 0$, $p \geq 0$ and the asymptotic conditions (7.9) and (7.10).

(c) Other technical hypotheses on the state equation which are physically comprensible as constraints on the adiabatic index of the fluid under question.

Let us now show how the techniques introduced in this paper are useful for the relativistic free-boundary problem.

As well as in the Newtonian case we study in a geometric and unified way both bounded and unbounded domains $\Omega$. All the classical approaches tackle only bounded regions of fluid. It is also immediate to see that the equations in (P3) are a particular case of the equations in (P1). Furthermore note that the coupling assumptions required in the proof of theorems 3.7 and 3.8 are very natural in the general relativistic setting and can be interpreted as consequences of the coupling (7.11).

However there is a fundamental assumption that (P3) does not satisfy: the analyticity of the metric tensor in the whole *M*. This fact, in principle, invalidates the application of corollary 3.9 to the problem (P3), however let us show that this problem can be overcome in two cases.

(i) A relativistic fluid model on a Euclidean space. This model does not consider the coupling appearing in (7.11) and it is used in some applications of interest [26]. Corollary 3.9 classifies the equilibrium shapes in ($R^n$, $d$) without any physical assumption. To our knowledge this result is novel in the literature. In general if we do not consider equation (7.11) corollary 3.9 applies and gives the equilibrium shapes in many different spaces (see section 5).

(ii) A relativistic fluid model on a conformally flat space considering the coupling (7.11). The metric tensor in appropriate local coordinates has the standard form $g_{ab} = e^{2f} d_{ab}$ where the function $f$ is analytic on the whole *M* except on $\partial\Omega$ where it is only $C_t^2$. The proof of theorem 3.6 is not applicable in this case, however theorem 3.6 is still correct for this kind of spaces as we sketch below. By assumption $f$ agrees fibrewise with *V* (this assumption is common in the literature [21]) and by ARP (theorem 4.5 holds in this case) both functions are analytically representable by the global analytic function *I*. Accordingly in certain open subset in a neighbouhood of a connected component of $\partial\Omega$ we can write $V=R(I)$ and $f=T(I)$. By following the notation used in the proof of theorems 3.6 and 4.5 we have that in $V_{out}$ equation (4.4) is substituted by

$$(R'(I)T'(I) + R''(I))\|\nabla_E I\|_E^2 + R'(I)\Delta_E I = 0 \qquad (7.12)$$

and hence according to (7.12) we have in $V_{out}$ that $\dfrac{\Delta_E I}{\|\nabla_E I\|_E^2} = c(I)$. By following the same procedure as in the proof of theorem 3.6 we reach identical conclusion: the partition induced by the potential function *V* is an equilibrium partition. The classification theorem 3.8 is therefore applicable and hence the equipotential hypersurfaces in ($R^n$, conformally flat) are geometrically trivial. This recovers in dimension 3 and bounded domain $\Omega$ a theorem of Lindblom [27] asserting that conformal flatness $\Rightarrow$ spherical symmetry. Note that our proof does not impose any physical or asymptotic assumption and is more general (valid for unbounded domains and arbitrary dimension *n*). In dimension 3 theorem 3.7 also gives information for conformally flat spaces with arbitrary topology, this being a new result in the literature.

The advantage of our approach is that one can study the equilibrium shapes in any space and obtain whether the partition is geometrically trivial or not. This technique is not adapted only to geometrically trivial partitions as the classical approaches are. For instance, the proof of Beig, Simon, Lindblom and Masood-ul-Alam in [24, 25] is adapted to prove that the domain Ω must be a perfect sphere under the physical assumptions (a), (b) and (c). On the contrary our approach can allow us to ascertain under what conditions the equilibrium shapes are perfect spheres and when they are not.

It would be desirable to be able to study the different metrics that can arise from coupling (7.11) and to generalize the concept of analytic representation of a metric tensor analogously to the conformally flat case discussed in (ii) above. Note that it is coherent and natural in this context to define the analytic representation of a metric and that (P3) must satisfy ARP on both $V$ and $g_{ab}$. This is due to the fact that in the problem (P3) the two global unknowns extending over the whole $M$ (across the boundary $\partial\Omega$) are the potential function $V$ and the metric tensor $g_{ab}$.

The following definition, with maybe some modifications, could work for metric tensors with isometries. It is immediate to show that it is consistent with the definition considered for conformally flat spaces.

**Definition 7.1.** The metric $g$ is analytically representable on the open set $U$ if there exists an analytic metric $\hat{g}$ whose Lie algebra $L$ of Killings is also a Lie algebra of Killings of $\hat{g}$.

Note that the dimension of $L$ is constant in $U$ and in fact $L$ is obtained via analytic continuation of the Killings in any open subset of $U$ [28]. For $U=M$ this implies that the isometries of the interior of the fluid are also isometries of the exterior when ARP on the metric tensor holds. This definition is inspired by the following one which is equivalent to definition 2.8 (note that the symmetries reconstruct the partition via Frobenius theorem).

**Definition 7.2.** $f$ is analytically representable on the open set $U$ if there exists an analytic function $I$ with the same symmetries as $f$. Recall that the vector field $\mathbf{y}$ is a symmetry of $f$ if and only if $\mathbf{y}(f) = 0$.

In the same spirit which connect definitions 2.8 and 7.2 one can change definition 7.1 to the following.

**Definition 7.3.** $g$ is analytically representable on the open set $U$ if there exists an analytic metric $\hat{g}$ such that $\mathbf{b}_U(g) = \mathbf{b}_U(\hat{g})$. The components of $\mathbf{b}_U(g)$ are defined as the transitivity sets of $L$ (two points $P$ and $Q$ are in the same transitivity set when $g(P)$ is isometric to $g(Q)$).

Regretfully, except in the conformally flat case, we are not aware of how to prove the relativistic analogous to theorems 3.6 and 4.5. If we could prove it then by applying the results of Kunzle [29] we could conclude that the solutions must be symmetrically spherical. Recall that in [29] Kunzle proves, under the strong assumption of $\|\nabla V\|$ being a function of $V$ and other physical conditions, that the spherical symmetry is unavoidable. Kunzle hypothesis on $\|\nabla V\|$ is a natural consequence of the partition induced by $V$ being an equilibrium partition.

To finish this subsection we would like to summary our proposal for tackling the relativistic free-boundary problem from the perspective developed in this paper:

(1) Formulating in a suitable manner the concept of analytic representation of a metric tensor, maybe in the line of definitions 7.1 and 7.3.

(2) Proving ARP (on both $V$ and $g_{ab}$) for problem (P3).

(3) Using a technique analogous to that employed in the proof of theorem 3.6 to obtain the same result: the partitions induced by the potential functions solutions to (P3) are equilibrium partitions.

(4) The problem is now reformulated: classifying equilibrium partitions of Riemannian manifolds. Our conjecture concerning this point is that theorem 3.6 holds when studying problem (P3) and that the equilibrium partition condition is a universal condition on the equilibrium equipotential hypersurfaces of a static fluid (Newtonian or relativistic) on any Riemannian manifold.

## Cosmological models.

Let us assume an inhomogeneous, static cosmological model on an unbounded Riemannian manifold $(M,g)$ that classically can be modeled by the problem (P2)'

$$\left.\begin{array}{l} \Delta V = r \quad \text{in } M \\ \nabla p + r \nabla V = 0 \quad \text{in } M \end{array}\right\} \qquad (7.13, 7.14)$$

and relativistically by the coupling (7.11) and the problem (P3)'

$$\left.\begin{array}{l} \Delta V = 4 p V (r + 3p) \quad \text{in } M \\ V \nabla p + (r + p) \nabla V = 0 \quad \text{in } M \end{array}\right\} \qquad (7.15, 7.16)$$

For the sake of simplicity we work under the assumption of $r$ and $p$ being real-valued analytic on the whole $M$ (and hence $V$ is also analytic by lemma 3.1).

The technique to prove theorem 3.6 is not valid for (P2)' and (P3)' since there is not external equation as (7.4) and (7.8). Anyway it is physically concevible that the following conjecture holds, as we next discuss:

**Conjecture 7.1.** If the solutions to the problems (P2)' and (P3)' satisfy certain asymptotic conditions guaranteeing that they tend to zero at infinity fast enough, then the partitions induced by them on $M$ are equilibrium partitions.

Physically, in the Newtonian case, one can reason as follows. Admit that $r$ and $p$ tend to zero at infinity in such a way that the total mass, the free energy ... are finite quantities. In this case, far enough of the central core, the functions $r$ and $p$ take small values compared with the values they take in the central core. Therefore, it is reasonable to approximate the problem (P2)' by the problem (P2) and hence theorem 3.6 holds. The key of this argument is that the strict mathematical infinite is physically closer.

# 8. Conclusion and open problems.

The technique we have used to understand the partitions induced by the solutions of (P1) on (*M*,*g*) has three different parts:

(1) Proving ARP which has been employed to get rid of the PDE system (P1) and to obtain the equilibrium function condition.

(2) Techniques of differential geometry of submanifolds to study each connected component of the fibres of the equilibrium functions on different spaces (*M*,*g*).

(3) Tecniques of real-valued analytic functions to globalize the local results obtained in the previous step.

The theorems obtained with the technique explained above have been succesfully applied to the classical problem of classifying the equilibrium shapes of Newtonian and relativistic static self-gravitating fluids.

Some interesting open problems follow:

(1) As a general problem it would be interesting to obtain more classification theorems of equilibrium partitions in other Riemannian spaces, at least in the 8 geometries modeling 3-dimensional spaces.

(2) Can all the Riemannian manifolds on which the equilibrium partitions are geometrically trivial or almost-trivial be classified? Can all the spaces on which the geometrical triviality and almost-triviality are equivalent concepts be classified?

(3) When an *n*-dimensional Riemannian manifold is flat and diffeomorphic to $R^n$ then the equilibrium partitions are geometrically trivial. If we change the topology of *M* but not the metric (it is still flat), how are the equilibrium partitions affected? How does the topology of the base manifold affect to the geometrical triviality? For flat manifolds, do a purely topological technique (of surgery type) exist to obtain the possible equilibrium partitions of (*M*,***d***) from the known equilibrium partitions of ($R^n$,***d***)?

(4) Are there Riemannian spaces on which there not exist equilibrium functions globally or locally?. If we fix the geometry of the base manifold, do topologies exist admitting that geometry for which there not exist equilibrium functions globally or locally?. If we fix the topology of the base space, do geometries exist admitting that topology for which there not exist equilibrium functions globally or locally?. Need all the equilibrium partitions be coupled with the metric *g* in some way in order to exist on (*M*,*g*)?.

(5) Since the fluid region $\Omega$ of the problems (P2) and (P3) extends over the manifold, do geometries or topologies exist in order that $\Omega$ not be simply connected?. What is the answer to this question in the flat case (*M*, $\delta$)?. If true we would have that global properties of the base manifold are inherited by a local mass of fluid and hence we would be able to determine the topology of the ambient space by studying the deformation retracts of this mass (the fluid would be twisted along the topology of *M*).

(6) We have applied succesfully the techniques of this paper to static and isolated self-gravitating fluids. Can the same techniques be applied to isolated self-gravitating fluids in evolution (for instance, rotating fluids)?. On the other hand, are there other physical systems (apart from self-gravitating fluids) for which the techniques developed in this paper are useful?. For example, in the problems of propagating interfaces (as ocean waves, burning flames, material boundaries and computer vision) [31] the most interesting properties are of geometrical type. Are the ideas of this paper applicable in that context?.

(7) A straightforward generalization of the concept of equilibrium function is the following:

**Definition 8.1.** Let $(M,g)$ be an $n$-dimensional Riemannian manifold and let $(f_1,...,f_m)$ be a set of $m < n$ real-valued analytic functions on $M$. Let us call $f \equiv (f_1,...,f_m)$, $\|\nabla f\|^2 = (\|\nabla f_1\|^2,...,\|\nabla f_m\|^2)$ and $\Delta f = (\Delta f_1,...,\Delta f_m)$. If $f$, $\|\nabla f\|^2$ and $\Delta f$ agree fibrewise on $M$ we say that $f$ is an equilibrium set of $m$ functions. The partition induced by $f$ is called an $m$-equilibrium partition of $M$.

Two open problems concerning definition 8.1 are the following: the first one is to obtain classification theorems analogous to those obtained in section 5 (where $m=1$) for example in the Euclidean $R^n$. And the other one is, do PDE problems (apart from (P1)) exist for which the solutions must satisfy the equilibrium condition defined in 8.1?. This generalization could be interesting because the equilibrium partitions of manifolds are very related to the topology and geometry of the manifold and hence the study of $m$-equilibrium partitions could provide useful tools for the study of the topology and geometry of manifolds (as new topological invariants).

(8) The reader must note that a partition of a manifold does not correspond only to a single function but rather to an equivalence class of functions satisfying the equivalence relation: $g \sim h$ iff $\boldsymbol{b}_M(g) = \boldsymbol{b}_M(h)$. However when we work to classify an equilibrium partition of a manifold we assume tacitly a representative of the class $f$ with which we work to do the computations. The equilibrium function condition is therefore a geometrical and topological condition rather than an analytical condition.

Taking into account the previous discussion the authors have the feeling that this kind of PDE problems, like (P1), restricting so much the topology and geometry of the fibres of their solutions should be possible to fit in a different and more topological and geometrical setting. In other words, could be possible to state a problem $\widetilde{(P)}$ in which the solutions were not functions but rather partitions of a manifold or equivalently their corresponding equivalence classes of continuous functions?. What we are suggesting is that the problem (P1) could be the statement in a PDE framework of another problem $\widetilde{(P1)}$ in a partition theory framework.

## 9. References.

[1] C. Cosner and K. Schmitt, Rocky Mt. J. Math. 18, 277-286 (1988). K.F. Pagani-Masciadri, J. Math. Anal. Appl. 174, 518-527 (1993). R. Gianni and R. Ricci, Adv. Math. Sci. Appl. 5, 557-567 (1995).

[2] C. Bar, Invent. Math. 138, 183-202 (1999). R. Hardt et al, J. Diff. Geom. 51, 359-373 (1999).

[3] P. Scott, Bull. London Math. Soc. 15, 401-487 (1983).

[4] R. Narasimhan: Introduction to the Theory of Analytic Spaces. Berlin: Springer-Verlag, 1966.

[5] A.B. Brown, Trans. Am. Math. Soc. 38, 379-394 (1935). W.F. Newns, Amer. Math. Month. 74, 911-920 (1967).

[6] M. Spivak: A Comprehensive Introduction to Differential Geometry (5 vols.). Berkeley: Publish or Perish, 1979.

[7] F. Gonzalez-Gascon, Phys. Lett. A 240, 147-150 (1998).

[8] A. Avez: Differential Calculus. Chichester: Wiley, 1986.

[9] D. Kinderlehrer, L. Nirenberg and J. Spruck, J. Analyse Math. 34, 86-119 (1978).

[10] C.B. Morrey and L. Nirenberg, Comm. Pure Appl. Math. 10, 271-290 (1957). C.B. Morrey: Multiple Integrals in the Calculus of Variations. Berlin: Springer, 1966.

[11] F. Broglia and A. Tognoli, Ann. Inst. Fourier, Grenoble 39, 611-632 (1989).

[12] H. Karcher: Riemannian comparison constructions. In Global Differential Geometry, Ed. S. S. Chern. MAA Studies in Mathematics, vol. 27, 1989.

[13] U. Lumiste: Submanifolds with parallel fundamental form. In Handbook of Differential Geometry, 779-864. Amsterdam: North-Holland, 2000.

[14] S. Helgason: Differential Geometry, Lie Groups and Symmetric Spaces. N. Y.: Ac. Press, 1978.

[15] M. Gromov, Rend. Sem. Mat. Fis. Milano 61, 9-123 (1991).

[16] S.G. Krantz and H.R. Parks, J. Diff. Eqs. 40, 116-120 (1981). R.L. Foote, Proc. Amer. Math. Soc. 92, 153-155 (1984).

[17] B. Segre, Rend. Acc. Naz. Lincei 27 (1938) 203-207.

[18] A.D. Alexandrov, Amer. Math. Soc. Trans. (2) 21, 412-416 (1962).

[19] R. Abraham, J.E. Marsden, T. Ratiu: Manifolds, Tensor Analysis and Applications. N. Y.: Springer, 1988.

[20] H. Poincare: Figures d'equilibre. Paris: Gauthier-Villars, 1902. R. Wavre: Figures Planetaries et Geodesie. Paris: Gauthier-Villars, 1932. L. Lichtenstein: Gleichgewichtsfiguren Rotiender Flüssigkeiten. Berlin: Springer, 1933. H. Lamb: Hydrodynamics. N. Y.: Dover, 1945.

[21] L. Lindblom, J. Math. Phys. 18, 2352-2355 (1977). L. Lindblom, Phil. Trans. R. Soc. Lond. A 340, 353-364 (1992).

[22] H. Muller zum Hagen, Proc. Camb. Phil. Soc. 67, 415-421 (1970).

[23] J.L. Synge: Relativity, the general theory. Amsterdam: North-Holland, 1966.

[24] R. Beig and W. Simon, Commun. Math. Phys. 144, 373-390 (1992).

[25] L. Lindblom and A.K.M. Masood-ul-Alam, Commun. Math. Phys. 162, 123-145 (1994).

[26] S.C. Noble and M.W. Choptuik: Collapse of relativistic fluids. Work in progress http://laplace.physics.ubc.ca/~scn/fluad (2002).

[27] L. Lindblom, J. Math. Phys. 21, 1455-1459 (1980).

[28] K. Nomizu, Ann. Math. 72, 105-120 (1960).

[29] H.P. Kunzle, Commun. Math. Phys. 20, 85-100 (1971).

[30] L. Lindblom, Astrophys. J. 208, 873-880 (1976).

[31] J.A. Sethian: Level Set Methods and Fast Marching Methods. Cambridge: Cambridge Univ. Press, 1999.